\documentclass[journal]{new-aiaa}
\usepackage[utf8]{inputenc}
\usepackage{textcomp}
\usepackage{graphicx}

\usepackage{amssymb}
\usepackage{amsmath}
\usepackage[version=4]{mhchem}
\usepackage{siunitx}
\usepackage{longtable,tabularx}
\usepackage{algorithm}
\usepackage{algorithmic}
\usepackage{booktabs}

\renewcommand{\Re}{\mathbb{R}}
\newcommand{\tp}{^{\mbox\tiny\mathsf{T}}}

\usepackage{tikz}
\usetikzlibrary{arrows.meta}
\let\oldincludegraphics\includegraphics
\renewcommand{\includegraphics}[2][]{%
	\scalebox{0.75}{\oldincludegraphics[#1]{#2}}}

\title{Nonlinear Thermal Modeling and Predictive Control for Spacecraft Optical Temperature Regulation}

\author{Marcus C. Klupar \footnote{Graduate Student, Department of Aerospace \& Mechanical Engineering.}, Daniel T.~Larsson\footnote{Assistant Professor, Department of Aerospace \& Mechanical Engineering.}, and Ewan S. Douglas\footnote{Associate Professor, Department of Astronomy and Aerospace \& Mechanical Engineering.}}
\affil{University of Arizona, Tucson AZ, 85721}

\begin{document}
\maketitle

\begin{abstract}
Maintaining a stable thermal environment is critical for spacecraft carrying precision optics, as temperature gradients and thermal expansion can alter the geometric alignment of optical components, compromise wavefront sensing, and degrade image quality even when active thermal systems provide real-time control.
In particular, the temperature of the optical payload must be controlled to sub-Kelvin tolerances about the desired setpoint, a requirement that is often difficult to achieve with conventional reactive control methods.
To this end, this paper develops a novel nonlinear thermal model for the optical system of a spacecraft, accounting for conduction, radiation, and the changing radiative environment induced by the spacecraft's orbit.
Using this model, we develop a model predictive control (MPC) methodology that enables closed-loop thermal regulation of the optical compartment under realistic orbital conditions.
Since only partial, noise-corrupted temperature measurements are available, an Unscented Kalman Filter is employed to estimate the full thermal state, yielding a certainty-equivalence MPC controller obtained via recursive online optimization.
The thermal model is validated against a high-fidelity thermal simulator, and the proposed approach is compared against conventional PID control, the prevailing method for on-orbit thermal regulation.
\end{abstract}

\section{Introduction}

\lettrine{T}{he} success of any spacecraft mission revolves around the reliability of its components. These components are strained through structural, thermal, and high intensity radiative environmental effects. To ensure component thermal reliability, it is important for components to stay within their designed temperature limits. Generally, thermal reliability is key for mission success, and thus engineers specifically design thermal systems to keep components safe over a range of operating conditions~\cite{speqtre, morgan2021optical, ashcraft2021versatile}.
Because of this, all spacecraft launched into space require some consideration of temperature regulation to ensure that key components do not leave their temperature operating range.
Approaches to thermal management can be broadly divided into two categories, active and passive regulation. Active regulation requires input power to cool or heat the spacecraft, while passive regulation involves the modification of natural heat dissipation through the choice of properties like emissivity~\cite{NASA}.
Spacecraft thermal control typically employs a combination of active and passive approaches, depending on mission requirements.

These requirements become especially demanding for spacecraft carrying precision optical payloads. To motivate our approach, we consider an archetypal 12U CubeSat in a Noon-Midnight orbit at 500 km carrying an optical payload as our spacecraft. To prevent unwanted thermal effects on the optics from sources like Solar, Earth IR, and Albedo radiation, active temperature regulation is needed to ensure the optical system stays within its design limits. 
In particular, maintaining wavefront sensing performance and image quality requires that the temperature of the optical payload be controlled to within sub-Kelvin tolerances of the desired setpoint~\cite{brooksPTC,gaudi2020habitable,knapp2020demonstrating,li2025design,song2025thermal}. These strict temperature operating ranges are needed to prevent thermal expansion across the optical system, which would in turn negatively impact optical performance~\cite{Brooks2022,kristwavefrontcontrol,brooksThermalControlTest}.

Oftentimes, active regulation comes in the form of surface heaters~\cite{brooksPTC,brooksThermalControlTest} controlled by a Proportional, Integral, and Derivative (PID) controller ~\cite{Brooks2022}.
A benefit of employing PID control is that the controller requires no explicit system model to perform corrections for the external thermal environment.
However, since PID controllers do not leverage a system model for prediction, they are inherently reactive, responding to disturbances only after they have occurred.
Consequently, classical PID-based thermal control provides limited tracking performance for time-varying disturbances, such as orbital heating, since it lacks an explicit model of future disturbance evolution required for disturbance rejection~\cite{moreno2020asymptotic}.

An alternative to PID is Model Predictive Control (MPC), in which a dynamical system model is leveraged to determine control inputs by solving an optimization problem over a specified time horizon.
In more detail, MPC is a recursive procedure that minimizes (or maximizes) a provided cost function over the set of allowable inputs at each time step to yield a closed-loop controller~\cite{MPC_engineering_perspective,MPC_past_present_future}, and uses the dynamical model to predict the impact of current control sequences on future temperature trajectories.
MPC offers several advantages over PID in our setting.
First, the control objective and system performance can be explicitly specified through a user-defined cost function and directly optimized, rather than tuned indirectly through PID gains.
Second, the dynamical (thermal) model underlying MPC provides a state-space formulation suitable for use alongside estimation methods such as the Unscented Kalman Filter (UKF) for validation and verification of system design requirements.
Third, since MPC is optimization-based, input constraints, such as maximum available heating power, can be enforced directly in the control design.
Finally, MPC naturally handles the joint control design of multiple (heater) inputs.

Although MPC has several advantages over PID-based control, a significant challenge to employing MPC is determining a representative dynamical (thermal) predictive model for designing the control inputs. Because MPC is a recursive online optimization problem, the computational efficiency of querying the system model for prediction is critical for real-time control. 
To address this challenge, existing MPC thermal control methods linearize the thermal model, which when paired with a quadratic performance measure yields a quadratic program that can be optimized efficiently~\cite{blet2002nonlinear}, at the cost of sacrificing nonlinear radiation effects~\cite{brooksPTC,chourasia2026reduced}.
In contrast, nonlinear thermal models are generally more computationally demanding, as the resulting control optimization requires nonlinear programming.
However, we observe that the thermal dynamics of our optical system are inherently nonlinear and evolve over relatively long time scales. This motivates an alternative approach: rather than linearizing the prediction model to reduce computational cost, we instead keep the nonlinear thermal model and use longer prediction time-step intervals. Because the nonlinear model provides accurate predictions over these larger intervals, the increased time between control updates provides additional available computation time to solve the more demanding nonlinear optimization while still satisfying real-time control requirements. By preserving the nonlinear radiation dynamics, we improve prediction fidelity over longer horizons and, ultimately, enhances control performance without sacrificing real-time implementation.

The development of a suitable thermal model gives rise to nonlinear state-space representations of the system, which are subsequently controlled by a recursive MPC algorithm.
However, implementing such a recursive procedure requires access to the full system state at each time step to initialize the optimization.
In practice, full state knowledge (i.e., the temperature at all locations within the optical system) is generally unavailable due to sensing limitations, and the sensors through which the control system observes the thermal environment are imperfect and noisy.
To overcome these challenges, an Unscented Kalman Filter (UKF) is employed to produce estimates of the full thermal state of the spacecraft from available measurements.

Our objective in this paper is therefore to develop a lightweight, high-fidelity thermal model for MPC-based predictive control, together with a control and estimation framework that enables efficient closed-loop regulation of an optical thermal environment when only partial, noise-corrupted measurements are available. 
Implementing such a recursive MPC procedure requires access to the full system state at each time step, which is generally unavailable in practice due to sensing limitations and
measurement noise.
Consequently, we implement an UKF to estimate the full thermal state from available measurements. 
We develop a novel semi-implicit nonlinear thermal model accompanied by an optimization algorithm for prediction over long horizons, building on thermal finite difference approaches~\cite{obrien1950study}. The semi-implicit formulation retains the nonlinear radiation model while achieving the computational efficiency required for MPC. 
This thermal model is then integrated into an predictive control framework operating in conjunction with a UKF to enable closed-loop control of the spacecrafts optics about a desired setpoint from noisy, partial temperature measurements.

The remainder of the paper is organized as follows.
Section~\ref{Thermal Modeling Section} derives the nonlinear spacecraft thermal model along with a semi-implicit numerical method for solving the governing thermal equations.
Section~\ref{Model Predictive Control Section} delineates the MPC procedure, which leverages the thermal model and processes noise-corrupted thermal measurements via an Unscented Kalman Filter to provide closed-loop thermal regulation of the optical system.
Section~\ref{Results Section} presents the performance and analysis of our approach in controlling a sixty-four node CubeSat model under realistic orbital conditions, followed by concluding remarks in Section~\ref{sec:conclusion}.
In the sequel, we denote the set of real numbers by $\mathbb{R}$, and, for any integer $n > 0$, the spaces of $n$-dimensional real-valued vectors and $n \times n$ real-valued matrices by $\mathbb{R}^n$ and $\mathbb{R}^{n\times n}$, respectively.

\section{Thermal Modeling}\label{Thermal Modeling Section}
 The development of our spacecraft finite-difference thermal model starts by defining the thermal properties. 
 In this paper, we assume all components of the spacecraft are made of Aluminum\footnote{Note that our approach is valid for any choice of material.} with density $\rho$, thermal conductivity $k$, specific heat $c_p$ and emissivity $\epsilon$.
 The control volume has volume $(V)$, average temperature $(T)$, and cross-sectional area $(A)$. 
 Due to the orbital environment, the spacecraft will only experience conduction and radiation as its modes of heat transfer.
Specifically, from the energy balance equation which describes the time rate of change of energy for a general system control volume, we have
 \begin{equation} \label{Continuous Energy Balance}
     \dot{E}_{st} = \dot{E}_{in} - \dot{E}_{out} + \dot{E}_{gen},
 \end{equation}
 where the energy storage of the system is described by $\dot{E}_{st}$, and can be thought of as a capacitance term. The energy in and out of the system is given by $\dot{E}_{in}$ and $\dot{E}_{out}$. 
 Finally, the internal energy generation is described by $\dot{E}_{gen}$, which will be our control variable in the sequel.  
 By discretizing Eq.~\eqref{Continuous Energy Balance} into control volumes and determining the energy entering and leaving a arbitrary control volume through conduction and radiation, we can derive a finite-difference thermal model, as follows. 
 First, in Section~\ref{Conduction Subsection}, we derive the energy entering, leaving, and being stored by an arbitrary control volume through conduction, before considering the energy entering or leaving a arbitrary control volume by radiation in Section~\ref{Radiation SubSection}. 
 We then combine the energy flows for conduction and radiation together to obtain an explicit thermal model in Section~\ref{Conduction and Radiation Subsection}. 
 Lastly, in Section~\ref{Semi-Implicit Method}, we convert the explicit thermal model into a more favorable semi-implicit representation. 
 The model is validated by comparing our semi-implicit thermal model to Thermal Desktop, which is a high-fidelity thermal solver.

\subsection{Energy Storage and Conduction} \label{Conduction Subsection}
\begin{figure}[htbp]
    \centering
    \includegraphics[width=0.5\textwidth]{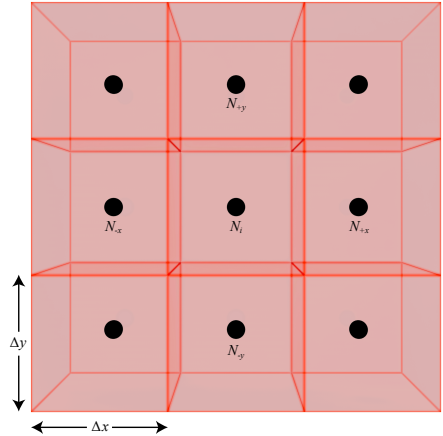}
    \caption{Generic mesh of the optical system. The mesh is comprised of control volumes (red lines), and nodes (black dots) that represent the systems physical properties. The spacecraft's structure (not shown) encapsulates the entire optical system. The structure and optical system are radiatively coupled as it is assumed the optical system mounts have low thermal conductivity.}
    \label{mesh}
\end{figure}

To derive our finite-difference thermal model to use in our MPC algorithm, we first discretize our spacecraft's optical system into \(n\) control volumes of equal dimensions, as shown in Fig.~\ref{mesh}. 
At the center of each control volume is a node which represents all the physical properties of its control volume (\(k,A,c_p,\rho,V,T\)).

With the mesh in Fig. \ref{mesh}, we can define a energy balance for the arbitrary node $i$ according to
\begin{equation}\label{Total Energy Balance}
    \dot{E}_{i,st} = \dot{E}_{i,in} - \dot{E}_{i,out} + \dot{E}_{i,gen}.
\end{equation}
Since the direction of heat flow is often unknown, it is common to assume that all the energy is entering the node of interest (\(\dot{E}_{i,st} = \dot{E}_{i,in} + \dot{E}_{i,gen}\)) \cite{IHT}.
For example, the total energy entering \(N_i\) from its surrounding nodes by means of conduction is,
\begin{equation}\label{Energy Entering}
    \dot{E}_{i,cond} = \dot{E}_{(+x) \rightarrow (i)} + \dot{E}_{(-x) \rightarrow (i)} + \dot{E}_{(+y) \rightarrow (i)} + \dot{E}_{(-y) \rightarrow (i)} + \dot{E}_{(+z) \rightarrow (i)} + \dot{E}_{(-z) \rightarrow (i)},
\end{equation}
with,
\begin{equation}
 \dot{E}_{(+x) \rightarrow (i)} = -kA_x\frac{\partial T}{\partial x} = kA_x\frac{T_{+x}^t - T_i^t}{\Delta x},
\end{equation}
which describes the rate at which energy moves from $N_{+x}$ to $N_i$.
Here we approximate the spatial derivative as a finite difference between the nodes \(N_i\) and \(N_{+x}\). A similar definition can also be found for the \(y\) and \(z\) directions.

We  follow a similar procedure for the energy storage term. 
Since the energy storage term in Eq.~\eqref{Total Energy Balance} depends on a time derivative, we discretize it in time~\cite{IHT} according to
\begin{equation}\label{Energy Storage}
    \dot{E}_{i,st} = \rho c_p V \frac{\partial T}{\partial t} = \rho c_p V \frac{T^{t+1}_i - T^t_i}{\Delta t},
\end{equation}
which describes the rate at which energy is being stored. 
A positive $\dot{E}_{st}$ indicates the control volume is heating while a negative value indicates cooling.
Relations~\eqref{Energy Entering} and~\eqref{Energy Storage} constitute the storage and conduction terms we set out to identify. 
We next derive the energy entering or leaving a control volumes surface through radiation.

\subsection{Energy Flow due to Radiation}\label{Radiation SubSection}

To develop our thermal model we must quantify the energy entering or leaving node \(N_i\) through radiation. 
To do so, we assume that the radiation is diffuse, gray, and uniform. This assumption is valid when the wavelength dependent surface properties remain relatively constant over the expected temperature range, which is true for our application since our spacecraft is expected to stay in the realm of 305 K. The spacecraft's optical system is sectioned into isothermal surfaces that match the discretizations made for conduction. The spacecraft structure is assumed to be a thin shell and thus requires only one surface and node. 
Each surface has surface properties such as emissivity \(\epsilon_i\), surface area \(A_i\), temperature \(T^t_i\), and a net radiation loss \(q^t_i\) \cite[pp. 218 - 220]{TRHT}.

\begin{figure}[h]
    \centering
    \includegraphics[width=0.8\textwidth]{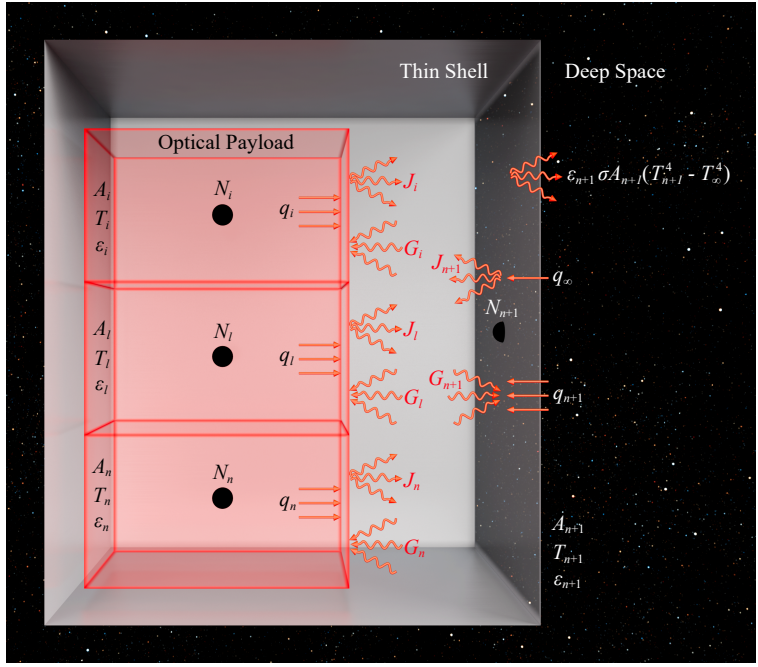}
    \caption{A gray, diffuse, and uniform radiosity enclosure is assumed for the spacecraft. Each node, denoted by the subscripts $i$ and $l$, has a exposed surface which has properties of surface area ($A$), temperature ($T$), and emissivity ($\epsilon$). These exposed surfaces reflect and emit ($J$) as well as absorb ($G$) radiation. In addition, each surface has a net radiation loss defined by $q_{li}$, while $q_{\infty}$ is the total absorbed radiation which is caused by the spacecraft's orbit (Solar, Albedo, and IR Radiation).}
    \label{Radiation Enclosure}
\end{figure}

The complex behavior of emission, absorption, and reflection can be modeled by observing the net radiation entering and leaving each surface \cite{TRHT}. 
Doing so results in the following system of equations
\begin{align} \label{Net Radiation Eq}
    \sum_{i=1}^{n+1} \left( \frac{\delta_{li}}{\epsilon_i} - F_{li} \frac{1-\epsilon_i}{\epsilon_i} \right)\frac{q^t_i}{A_i} = \sum_{i=1}^{n+1}\sigma(\delta_{li} - F_{li})(T_i^4)^t, \qquad l = 1,2,..,n+1 ,
\end{align}
which describe the net energy transfer from the surface of $N_l$ to the surrounding surfaces $N_i$ through radiation.
Since there are \(n\) surfaces for the optical system and one surface for the structure, the summation is done over the \(n+1\) total surfaces. 
The subscript \(li\) defines the surfaces that interact, for example \(F_{li}\) is the view factor from surface \(l\) to surface \(i\), which describes the percent of radiation emitted from surface $l$ that is incident on surface $i$. It is important to note that in order for this system of equations to be solved, each surface must have either a temperature (\(T^t_i\)) or flux boundary condition (\(q^t_i\)) defined for it, which is currently not met as only the temperatures \(T^t_1\) through \(T^t_n\) are assumed known, leaving surface \(n+1\) currently without a known temperature or a flux boundary condition as it is a function of the spacecrafts external radiation environment.

Consequently, Eq.~\eqref{Net Radiation Eq} has \(n+1\) equations with \(n+2\) unknowns ($q^t_1, ...,q^t_n,q^t_{n+1},T^t_{n+1}$). In order to solve this system of equations we will need an additional equation, which we obtain by considering the radiation outside the spacecraft.

\begin{figure}[h]
    \centering
    \includegraphics[width=0.8\textwidth]{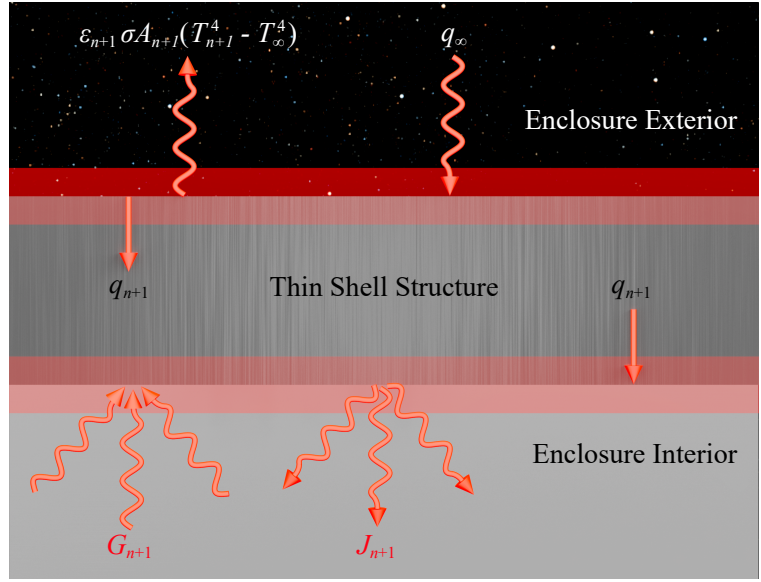}
    \caption{The spacecraft structure, which is assumed to be a thin shell, participates in a radiation exchange with the interior surfaces and the exterior deep space. The thin shell structure receives (\(G_{n+1}\)) as well as reflects and emits (\(J_{n+1}\)) radiation from the spacecraft interior. Additionally, the thin shell structure emits radiation to deep space (\(\epsilon \sigma A (T_{n+1}^4 - T_\infty^4)\)), while also absorbing radiation from the sun and the earth (\(q_\infty\)). \(q_\infty\) is a function of the spacecraft orbit.}
    \label{External Radiation}
\end{figure}

The energy balance for the exterior of the thin shell structure is described as,
\begin{align} \label{structure energy balance}
    0 = \dot{E}_{in} - \dot{E}_{out} &= q^t_{\infty} - (q^t_{n+1} + \epsilon_{n+1} \sigma A_{n+1} ((T_{n+1}^4)^t - T_\infty^4)), \\ \nonumber
    q^t_{n+1} &=  q^t_{\infty} - \epsilon_{n+1} \sigma A_{n+1} ((T_{n+1}^4)^t - T_\infty^4),
\end{align}
where $q_\infty^t$ is a function of the spacecrafts orientation and orbital position and represents the absorbed radiation from the Sun, Earth IR, and Albedo at a given time $t$. $T_\infty$ is the temperature of deep space which is typically defined as $2.73$K.
Combining Eq.~\eqref{Net Radiation Eq} and Eq.~\eqref{structure energy balance} leads to \(n+2\) equations with \(n+2\) unknowns.
These two equations can  be written together as the system
\begin{equation} \label{Total System of Eq}
\begin{bmatrix}
    L_{1,1}  & ... & L_{1,n+1} \\
    \vdots & \ddots & \vdots\\
    L_{n+1,1} & \dots  & L_{n+1,n+1} \\
    0 &  \dots  & 1
\end{bmatrix}
    \begin{bmatrix}
        q^t_1\\ \vdots\\ q^t_{n+1}
    \end{bmatrix}
    = 
 \begin{bmatrix}
    R_{1,1}  & ... & R_{1,n+1} \\
    \vdots & \ddots & \vdots\\
    R_{n+1,1} & \dots  & R_{n+1,n+1} \\
    0 &  \dots  & R_s
\end{bmatrix}
 \begin{bmatrix}
      (T_1^4)^t\\ \vdots\\ (T_{n+1}^4)^t
 \end{bmatrix}
 + 
 \begin{bmatrix}
     0\\ \vdots \\0 \\ R_\infty
 \end{bmatrix},
\end{equation}
where \(L_{li} = (\frac{\delta_{li}}{\epsilon_i} - F_{li}\frac{1-\epsilon_i}{\epsilon_i})\frac{1}{A_i}\), \(R_{li} = (\delta_{li} - F_{li})\sigma \), \(R_s = -\epsilon_{n+1}\sigma A_{n+1}\), and \(R_\infty = q^t_\infty -\epsilon_{n+1}\sigma A_{n+1} T_\infty^4 \).

Solving Eq.~\eqref{Total System of Eq} for \(q^t\big((T^t_1)^4,...,(T^t_{n+1})^4\big) = [q^t_1\big((T^t_1)^4,...,(T^t_{n+1})^4\big), \dots, q^t_{n+1}\big((T^t_1)^4,...,(T^t_{n+1})^4\big)]^{\tp} \), where $(\cdot)^{\tp}$ denotes the transpose operation, yields the net radiation loss from each of the \(n+1\) surfaces at time \(t\).
The value \(q^t\big((T^t_1)^4,...,(T^t_{n+1})^4\big)\) represents a net radiative loss since we assumed that the direction of \(q^t_i\big((T^t_1)^4,...,(T^t_{n+1})^4\big)\) to be leaving the control volume as seen in Fig. \ref{Radiation Enclosure}. 
Therefore, the energy entering node \(N_i\) through radiation is denoted as,
\begin{equation} \label{Energy Entering Rad}
    \dot{E}_{i,rad} = -q^t_i\big((T^t_1)^4,...,(T^t_{n+1})^4\big).
\end{equation}
It should be noted that the net radiative energy loss from any arbitrary node, \(q^t_i\big((T^t_1)^4,...,(T^t_{n+1})^4\big)\), is a nonlinear function of the temperatures \(T^t_1\) through \(T^t_{n+1}\), but the shorthand \(q_i^t \equiv q_i^t\big((T^t_1)^4,...,(T^t_{n+1})\big)\) will be used for ease of notation.
To complete the thermal model we must combine the energy entering node \(N_i\) by conduction and radiation, as discussed next.

\subsection{Total Net Energy Flow: Storage, Conduction, and Radiation Combined}\label{Conduction and Radiation Subsection}

We previously found in Section~\ref{Conduction Subsection} that the energy entering node \(N_i\) through conduction is given by Eq.~\eqref{Energy Entering}. 
We also found in Section \ref{Radiation SubSection} that the energy entering \(N_i\) through radiation is given by Eq.~\eqref{Energy Entering Rad}. By adding these two equation together we get the total net energy flow $\dot{E}_{i,in} = \dot{E}_{i,cond} + \dot{E}_{i,rad}$ for node \(N_i\) as
\begin{equation}\label{Total Energy Entering}
    \dot{E}_{i,in} = \dot{E}_{(+x) \rightarrow (i)} + \dot{E}_{(-x) \rightarrow (i)} + \dot{E}_{(+y) \rightarrow (i)} + \dot{E}_{(-y) \rightarrow (i)}+ \dot{E}_{(+z) \rightarrow (i)} + \dot{E}_{(-z) \rightarrow (i)} + \dot{E}_{i,rad}.
\end{equation}
Substituting Eq.~\eqref{Total Energy Entering} and Eq.~\eqref{Energy Storage} into our total energy balance equation given by Eq.~\eqref{Total Energy Balance}, we obtain
\begin{align*} 
    &\rho c_pV \frac{T^{t+1}_i - T_i^t}{\Delta t} = \dot{E}_{(+x) \rightarrow (i)} + \dot{E}_{(-x) \rightarrow (i)} + \dot{E}_{(+y) \rightarrow (i)} + \dot{E}_{(-y) \rightarrow (i)}+ \dot{E}_{(+z) \rightarrow (i)} + \dot{E}_{(-z) \rightarrow (i)} +\dot{E}_{i,rad} + \dot{E}_{i,gen}.
\end{align*}
Solving for the future temperature of node \(N_i\) gives,
\begin{equation} \label{3D energy balance for Ni}
    T_i^{t+1} = \frac{k \Delta t}{\rho c_p V} \bigg(A_x\frac{T^t_{+x} + T^t_{-x} - 2T_i^t}{\Delta x} + A_y\frac{T^t_{+y} + T^t_{-y} - 2T_i^t}{\Delta y} + A_z\frac{T^t_{+z} + T^t_{-z} - 2T_i^t}{\Delta z}\bigg) - \frac{\Delta t}{\rho c_p V}q^t_i + \frac{\Delta t}{\rho c_p V} \dot{E}_{i,gen} + T^t_i.
\end{equation}
Eq.~\eqref{3D energy balance for Ni} models how the temperature of node \(N_i\) changes with time. From Eq.~\eqref{3D energy balance for Ni}, we also see that the temperature of node \(N_i\) depends on its surrounding nodes. Therefore, we can write Eq.~\eqref{3D energy balance for Ni} over all of the \(n\) nodes comprising the optical system according to
\begin{equation} \label{explicit}
    [T]^{t+1} = M_{cond}[T]^t - \frac{\Delta t}{\rho c_p V}[q]^t + \frac{\Delta t}{\rho c_p V} [\dot{E}_{gen}]^t,
\end{equation}
where $M_{cond}$ is a conduction coefficient matrix of size \(n \times n\). 
The bracket notation used here indicates a column vector of size $n \times 1$ where each row $i$ corresponds to node $i$ (e.g., row 3 of $[T]^t$ is the temperature of node $N_3$ at time $t$). 
The vector \([\dot{E}_{gen}]^t\) is the internal heat generation term at time $t$, and is used to control the temperature of the spacecraft.
It is important to note that although $[\dot{E}_{gen}]^t$ is an $n\times 1$ vector in Eq.~\eqref{explicit}, it need not be the case that every node have the ability to add energy to the system.
Further details will be provided in Sections~\ref{Model Predictive Control Section} and~\ref{Results Section}.

The relation given in Eq.~\eqref{explicit} is a discrete-time, explicit thermal model that can be used to predict future temperatures in an MPC approach.
However, a significant drawback of an explicit model for temperature prediction is the inverse relationship between the time-step size $\Delta t$ and the number of thermal model nodes required to maintain numerical stability~\cite{IHT,keppens1999implicit}.
In other words, increasing the spatial resolution of the model (i.e., adding more nodes) requires $\Delta t$ to decrease in order to maintain numerical stability.
Consequently, for a given prediction horizon, the number of prediction points required in the MPC optimization, and thus the dimension of the associated optimization problem, grows quickly and limits the predictive capability of the approach.
Therefore, in the next section, we develop a semi-implicit method that circumvents these numerical stability challenges, allowing our approach to maintain sufficiently long prediction horizons while keeping the MPC optimization problem tractable.

\subsection{Semi-Implicit Method} \label{Semi-Implicit Method}

Inspired by the Crank-Nicolson method which averages an explicit and implicit relation to improve the numerical stability \cite{abhulimen2019modified}, we add the explicit Eq.~\eqref{explicit} to the following semi-implicit equation,
\begin{equation}\label{semi-implicit1}
    [T]^{t+1} = M_{cond}[T]^{t+1} - \frac{\Delta t}{\rho c_p V}[q]^t  + \frac{\Delta t}{\rho c_p V} [\dot{E}_{gen}]^t.
\end{equation}
The semi-implicit relation in Eq.~\eqref{semi-implicit1} is obtained by evaluating the conduction terms at time \(t+1\) instead of time \(t\), thereby employing a forward difference instead of a backward difference to approximate the derivatives.
The summation of Eq.~\eqref{explicit} and Eq.~\eqref{semi-implicit1} results in the semi-implicit relation
\begin{equation} \label{semi-implicit}
    [T]^{t+1} = \frac{1}{2} \bigg(M_{cond}[T]^{t+1} + M_{cond}[T]^{t} \bigg)  - \frac{\Delta t}{\rho c_p V}[q]^t  + \frac{\Delta t}{\rho c_p V} [\dot{E}_{gen}]^t,
\end{equation}
which is semi-implicit since the future temperature of the spacecraft depends on the current and future conduction values as well as current radiation and internal heating values. 
Solving Eq.~\eqref{semi-implicit} for \([T]^{t+1}\) leads to
\begin{equation}\label{thermal model}
    [T]^{t+1} = \big(I - \frac{1}{2}M_{cond}\big)^{-1} \bigg(\frac{1}{2}M_{cond}[T]^t  - \frac{\Delta t}{\rho c_p V}[q]^t  + \frac{\Delta t}{\rho c_p V} [\dot{E}_{gen}]^t \bigg),
\end{equation}
where $I$ is the $n \times n$ identity matrix.
The nonlinear, semi-implicit, thermal model described by Eq.~\eqref{thermal model} has the benefit of maintaining numerical stability for large values of \(\Delta t\), making it an excellent model to be used in our Model Predictive Control algorithm. 
The performance of the MPC algorithm improves the farther into the future it can predict, making the prediction horizon a key factor in control performance. For a given prediction horizon, the explicit model requires significantly more time steps than the semi-implicit model to maintain numerical stability, resulting in a substantially higher computational burden for the online optimization.

\subsection{Thermal Model Results and Validation} \label{Thermal Model Results}
In order to validate our thermal model given by Eq.~\eqref{thermal model}, we compare it to one generated by the high-fidelity thermal solver Thermal Desktop. 
In both models (Eq.~\eqref{thermal model} and Thermal Desktop), we considered a noon-midnight orbit with the spacecrafts z-axis always pointing toward the center of the Earth. 
The orbit parameters are, inclination = 78.8$^\circ$, Right Ascension of ascending node = 0$^\circ$, argument of periapsis = 270$^\circ$, eccentricity = 0, period = 5677 s, and an altitude of 500 km. Additionally, with no loss of generality, we realize the spacecraft as a archetypal 12U CubeSat carrying a optical payload to showcase our methods without a complicated spacecraft design.

\begin{figure}[htbp]
    \centering
    \includegraphics[width=0.8\textwidth]{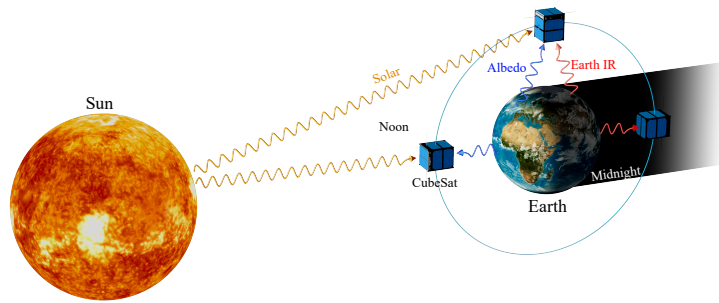}
    \caption{Simple schematic of a noon–midnight orbit showing the CubeSat, circular orbit, and radiation sources.}
    \label{fig:noon-midnight-orbit}
\end{figure}

\begin{figure}[ht!]
\centering
\includegraphics[width=.8\textwidth]{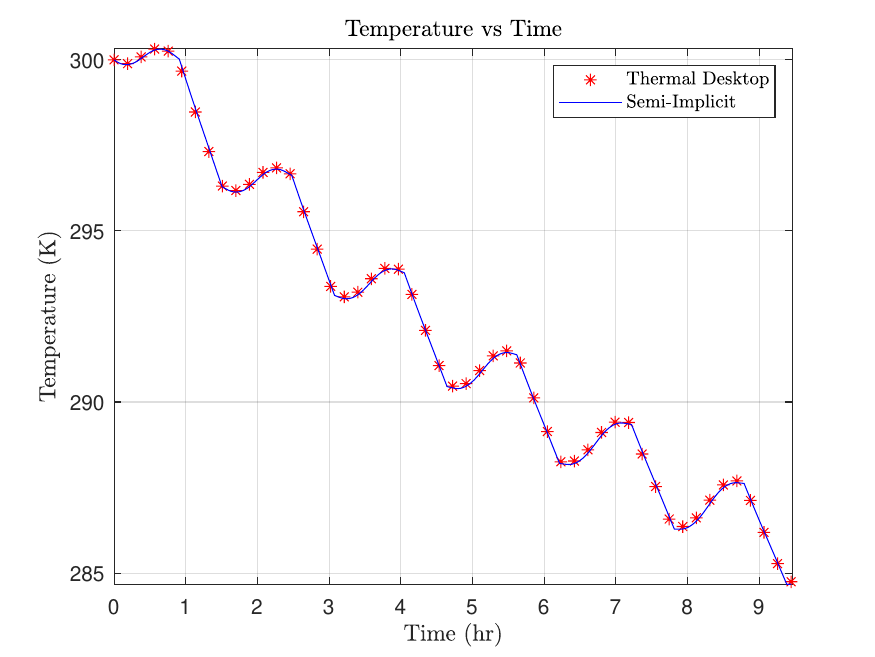}
\caption{Unactuated (free-response) temperature response for the CubeSats optical system while in a noon-midnight orbit with a initial condition of 300 \textmd{K}. The oscillatory behavior is due to the CubeSat being eclipsed by the earth  throughout its orbit. The red stars (*) represent the Thermal Desktop results, while the blue line represents Eq.~\eqref{thermal model}.}
\label{Free Response}
\end{figure}

Since the CubeSat is pointing down toward the Earth, the CubeSat's projected area relative to the sun will change with time. This results in a variable incident external heating. 
By using a single node to represent the CubeSat's optical system in both Eq.~\eqref{thermal model} and Thermal Desktop and by setting the internal heat generation term \(\dot{E}^t_{gen}\) to zero, we can compare the  response  of the two models as shown in Fig.~\ref{Free Response}. 
It is important to note that a single node representation for the optical payload is sufficient as it represents the average temperature of the entire optics. Increasing the number of nodes improves the thermal resolution by using more control volumes, but will not change the average temperature of the entire optics in a significant way.
The CubeSat is assigned an initial temperature of \(T^{t=0} = 300\) K, and deep space is given a temperature of \(T_\infty = 2.73\) K.

Fig.\ref{Free Response} shows a strong agreement between Eq.~\eqref{thermal model} and Thermal Desktop which validates the approach taken. The oscillatory behavior seen in Fig.\eqref{Free Response} is due to the CubeSat being eclipsed by the Earth in its noon-midnight orbit.

\section{Model Predictive Control \& Kalman Filtering} \label{Model Predictive Control Section}
Our motivation is to precisely control the temperature of the optical system to avoid effects such as thermal expansion that, if left unchecked, will deteriorate the optical performance. We can accomplish this goal by developing a model predictive control (MPC) algorithm that utilizes the CubeSat's thermal model to determine suitable heating inputs. The thermal model Eq.~\eqref{thermal model} is used to predict the future temperature of the CubeSat so that a sequence of heating inputs can be determined to maintain the optical system at its desired temperature.

In our system Eq.~\eqref{thermal model}, the internal heat generation term $[\dot{E}_{gen}]^{t}$ with units of Watts is the control input at each discrete time instance $t$.
To design the input $[\dot{E}_{gen}]^{t}$ using model predictive control, we will, at each time $t$, consider optimizing over input sequences of the form $[\dot{E}_{gen}]^{0|t},[\dot{E}_{gen}]^{1|t},\ldots,[\dot{E}_{gen}]^{h|t}$ that stretch $h$ time-steps into the future from time $t$ in order to design a controller that is reactive to the future temperature of the system resulting from the application of the input sequence.\footnote{Here, $[X]^{l|t}$ denotes the predicted value of $[X]$ at time $t+l$, while $[X]^t$ denotes its true value at time $t$.}
To evaluate the performance of each input sequence, we will minimize a quadratic cost function that utilizes the dynamical model Eq.~\eqref{thermal model} to predict the impact on the future temperature sequence $[T]^{0|t},\ldots,[T]^{h|t}$ as a result of the input sequence $[\dot{E}_{gen}]^{0|t},[\dot{E}_{gen}]^{1|t},\ldots,[\dot{E}_{gen}]^{h|t}$.
Once the sequence $[\dot{E}_{gen}]^{0|t},[\dot{E}_{gen}]^{2|t},\ldots,[\dot{E}_{gen}]^{h|t}$ is obtained, we apply the first input to our system so that $[\dot E_{gen}]^t = [\dot{E}_{gen}]^{0|t}$, then propagate the system forward in time according to Eq.~\eqref{thermal model}.
The process is then repeated to determine the next sequence of optimal inputs in a receding-horizon fashion.
The algorithm is shown in Algorithm~\ref{MPC Algorithm} \cite{MPC_engineering_perspective,MPC_past_present_future}.

More formally, the optimization problem we consider is given by
\begin{align} \label{cost function}
   \min_{[\dot{E}_{gen}]^{0|t},\ldots,[\dot{E}_{gen}]^{h|t}} \sum_{l=0}^h & \Bigr(\bigr( [T]^{l|t
   }-[T_{ref}] \bigl)\tp Q \bigr([T]^{l|t}-[T_{ref}]\bigl) \nonumber \\ 
   & \quad + ([\dot{E}_{gen}]^{l|t})\tp   R([\dot{E}_{gen}]^{l|t}) \nonumber \\
   & \quad +  ([\dot{E}]_{gen}^{l|t} - [\dot{E}_{gen}]^{l-1|t})\tp R_{\Delta}([\dot{E}_{gen}]^{l|t} - [\dot{E}_{gen}]^{l-1|t}) \Big),
\end{align}
subject to the dynamical constraint Eq.~\eqref{thermal model} and the input constraint
\begin{align}
    [lb] \leq[\dot{E}_{gen}]^{l|t} \leq [ub], ~ \quad l = 0,1, 2, \ldots,h,  \label{eq:MPCineqCons1}
\end{align}
where $[\dot{E}_{gen}]^{-1|t} \equiv [\dot{E}_{gen}]^{t-1}$, \([T]^{l|t} \in \Re^{n\times 1}\) is the temperature predicted by the thermal model \(l\) time steps ahead of the current time \(t\) obtained by forward-simulation of $[\dot{E}_{gen}]^{0|t},\ldots,[\dot{E}_{gen}]^{l-1|t}$ through the dynamics model Eq.~\eqref{thermal model}, and the inequality constraint~\eqref{eq:MPCineqCons1} is enforced component-wise.
The vector \([T_{ref}] \in \Re^{n\times 1}\) is the reference temperature for which the optical system is to be regulated about, and \([\dot{E}_{gen}]^{l|t} \in \Re^{n\times 1}\) is the predicted internal heating \(l\) time steps ahead of the current time \(t\). If only a subset of $[\dot{E}_{gen}]^{l|t}$ is used for heating, then the lower and upper bounds of the input constraints can be set to zero for the nodes that do not allow heating inputs.
The vectors $[lb]\in\Re^n$ and $[ub]\in \Re^n$ are the lower and upper bound constraints on the input over the prediction horizon, and \(Q \in \Re^{n\times n}\), \(R\in \Re^{n \times n}\), and $ R_\Delta\in \Re^{n \times n}$ are positive semi-definite weighting matrices penalizing the state, input, and input rate of change, respectively, the entries of which are used to adjust the performance of the algorithm. 
Minimizing the cost function described by Eq.~\eqref{cost function} results in a sequence $[\dot{E}_{gen}]^{0|t},\ldots,[\dot{E}_{gen}]^{h|t}$ specifying the input to the thermal system in the CubeSat at each time instance $l \in \{0,\ldots,h\}$ when the MPC predictive dynamic process is  initialized with the state $[T]^{t}$.
The true value of the applied input at time $t$ is then $[\dot E_{gen}]^t = [\dot{E}_{gen}]^{0|t}$, and the process is repeated at time-step $t+1$.

The control design process discussed above is implementable whenever the controller has access to the full state $[T]^t\in\Re^{n\times 1}$ of the thermal system.
However, in many real-world applications, the state, in this case the temperature vector $[T]^t\in\Re^{n\times 1}$, cannot be fully observed, as temperature is generally measured only at specific locations within the optics compartment, such as near extremities.
Moreover, even in cases where an observation of some or all of $[T]^t$ is obtainable, the measurement contains noise attributable to factors such as the conversion from analog to digital signals, sensor quality and condition, or environmental disturbances~\cite{fahey2000noise}. 
Consequently, a state-estimation process must be introduced in order to generate an estimate $[\hat T]^{t} \in \Re^{n \times 1}$ of $[T]^t \in \Re^{n\times 1}$ at each time $t$.

To this end, we implement an Unscented Kalman Filter (UKF) to generate an estimate $[\hat T]^t$ of the state $[T]^t$ from noisy measurements. 
The UKF is a state-estimation approach that extends ideas from the Kalman filter, originally developed for linear systems, to nonlinear systems such as our temperature control problem. 
To generate a state estimate, the UKF employs the Unscented Transform~\cite{wan2000unscented} to approximate distributions over the state $[T]^t$ using Gaussian distributions~\cite{wan2001unscented}. 
The UFK has been ubiquitously employed for state-estimation in nonlinear systems~\cite{ko2007gp,kandepu2008applying} and, importantly, does not require the dynamical system to be linearized through the computation of Jacobian's as is the case for Extended Kalman Filtering (EKF)~\cite{wan2000unscented}.
Since the UKF does not require linearization, it is able to capture system nonlinearities to a greater degree as compared with the EKF~\cite{wan2001unscented}.

In our setting, the dynamics of the true, unobserved, state are $[T]^{t+1} = f([T]^t,[\dot{E}_{gen}]^t,[w]^t)$ where the nonlinear function $f:\Re^n \times \Re^n \times \Re^n \to \Re^n$ is given by
\begin{equation} \label{noise thermal model}
  f([T]^t,[\dot{E}_{gen}]^t,[w]^t) = \big(I - \frac{1}{2}M_{cond}\big)^{-1} \bigg(\frac{1}{2}M_{cond}[T]^t  - \frac{\Delta t}{\rho c_p V}[q]^t  + \frac{\Delta t}{\rho c_p V} [\dot{E}_{gen}]^t \bigg) + [w]^t,  
\end{equation}
which is the same as the right-hand side Eq.~\eqref{thermal model} except for the addition of independent and identically distributed (iid) additive Gaussian-distributed process noise $[w]^t \sim \mathcal{N}([0], R^w)$ with mean $[0]\in\Re^n$ and covariance $R^w\in\Re^{n\times n}$ modeling disturbances or discrepancies in the developed thermal model. 
At each time $t$, the system generates an observation $[T_{obs}]^t \in \Re^{m}$ via thermal sensors attached to the optical payload according to
\begin{equation} \label{observation eq}
    [T_{obs}]^t = H[T]^t + [v]^t, 
\end{equation}
where $H \in \Re^{m \times n}$ is a matrix determining which of the $n$ thermal states are observed, and $[v]^t \sim \mathcal N ([0],R^v)$ is iid sensor noise disturbances distributed according to a zero-mean Gaussian distribution with covariance $R^v\in\Re^{m\times m}$. 
Since the controller does not have access to the full state $[T]^t$ at each time $t$, it instead must determine a suitable input using the information state $[\mathcal{I}]^t = \{[T_{obs}]^0,\ldots,[T_{obs}]^{t},[\dot{E}_{gen}]^0,\ldots,[\dot{E}_{gen}]^{t-1}\}$ comprised of past inputs and observations.
The objective of the UKF is to utilize the information state as well as knowledge of the system dynamics Eq.~\eqref{noise thermal model} and observation process Eq.~\eqref{observation eq} to generate a state estimate $[\hat T]^{t} \in \Re^n$ at time $t$.

To accomplish this goal, the UKF proceeds as follows. 
Assume that at time $t-1$, the UKF has generated the belief distribution $[T]^{t-1} \sim \mathcal {N}([\bar T]^{t-1},P^{t-1})$ with mean $[\bar T]^{t-1} \in \Re^n$ and covariance $P^{t-1} \in \Re^{n\times n}$ using $[\mathcal{I}]^{t-1}$.
At time $t$, a new measurement $[T_{obs}]^t$ and input $[\dot E_{gen}]^{t-1}$ become available, allowing the following two-step procedure to update the distribution over $[T]^{t-1}$ to one over $[T]^t$.
Note that $[\mathcal{I}]^t = [\mathcal{I}]^{t-1} \cup \{[T_{obs}]^t,[\dot{E}_{gen}]^{t-1}\}$ and that information regarding $[\mathcal{I}]^{t-1}$ has been previously used to generate $[\bar T]^{t-1}$ and $P^{t-1}$.

In the first step, the UKF uses the dynamic model Eq.~\eqref{thermal model} to propagate the sigma points $[\mathcal{X}]_0,\ldots,[\mathcal{X}]_{2n}$ in order to provide an estimate of the belief distribution after the application of $[\dot{E}_{gen}]^{t-1}$ but before the incorporation of the measurement $[T_{obs}]^t$. 
To this end, the sigma points are chosen according to 
\begin{align}
    [\mathcal{X}^{t-1}]_{0} &= [\bar{T}]^{t-1}, \nonumber\\
    [\mathcal{X}^{t-1}]_{i} & = [\bar{T}]^{t-1} + \big[\sqrt{(n+\lambda)P^{t-1}}\big]_i, \quad  i = 1,\ldots ,n, \\
    [\mathcal{X}^{t-1}]_{i} & = [\bar{T}]^{t-1} - \big[\sqrt{(n+\lambda)P^{t-1}}\big]_{i-n}, \quad i = n+1,\ldots,2n, \nonumber
\end{align}
where $\lambda = \alpha^2(2n+3)-n$, and $\alpha$ is a constant that determines the spread of the sigma points around the mean and is typically small (i.e., $1 \leq \alpha \leq 10^{-4}$)~\cite{wan2001unscented}. The subscript $i$ refers to the $i^\text{th}$ column of the matrix (i.e, $[\mathcal{X}^{t-1}]_i$ is the $i^{\text{th}}$ column of the matrix $\mathcal{X}^{t-1}$).
The sigma points are then passed through the nonlinear dynamics Eq.~\eqref{thermal model} and are used to estimate the mean and covariance of the belief distribution before the measurement update according to,
\begin{align}
    \mathcal{X}_*^{t|t-1} &= f(\mathcal{X}^{t-1}, [\dot{E}_{gen}]^{t-1},[0]), \quad i=0,...,2n.
\end{align}
The predicted mean and covariance of the belief distribution for the optics temperature before the inclusion of the measurements is then given by
\begin{align}
    [\bar{T}]^{t}_{(-)} &\approx \sum_{i=0}^{2n}\omega_i^{m} [\mathcal{X}_{*}^{t|t-1}]_i \\
    P^{t}_{(-)} &\approx \sum_{i=0}^{2n}\omega_i^{c} ([\mathcal{X}_{*}^{t|t-1}]_i - [\bar{T}]^{t}_{(-)})([\mathcal{X}_{*}^{t|t-1}]_i - [\bar{T}]^{t}_{(-)})\tp + R^w,
\end{align}
respectively, and where the weights are defined as
\begin{align}
    \omega_0^{m} &= \frac{\lambda}{n + \lambda}, \\
    \omega_0^{c} &= \frac{\lambda}{n + \lambda} + 1-\alpha^2+\beta, \\
    \omega_i^{m} &=\omega_i^{c} = \frac{1}{2(n+\lambda)}, \quad\quad i = 1,\ldots,2n,
\end{align}
with $\beta = 2$ \cite{wan2001unscented}.

The second step of the UKF then works to generate $[\bar T]^t$ and $P^t$ from $[\bar T]^{t}_{(-)}$ and $P^{t}_{(-)}$ by incorporating the measurement $[T_{obs}]^t$ in a similar procedure to that described above, but using the measurement model and measurement noise in place of the system dynamics and process noise, respectively. 
Note that, since the measurement model Eq.~\eqref{observation eq} is linear, the sigma-point propagation step is unnecessary and the standard Kalman measurement update is used directly.
The UKF equations and procedure is summarized in Table~\ref{UKF Table}.
In Table \ref{UKF Table}, $\gamma = \sqrt{n+\lambda}$, $R^w$ is the process-noise covariance, $R^v$ is the measurement-noise covariance, and $[T_{\text{obs}}]^{t}$ is the observation at time $t$. 
Once the distribution $[T]^t \sim \mathcal{N}([\bar T]^t,P^t)$ over the optics compartment temperature at time $t$ is known, an estimate is generated according to $[\hat T]^t = [\bar T]^t$.
For more details regarding the UKF its properties, the interested reader is directed to Wan and Van Der Merwe~\cite{wan2001unscented,wan2000unscented}.

\begin{table}[ht!]
\centering 
\renewcommand{\arraystretch}{1.6}
\begin{tabular}{p{0.32\linewidth} p{0.63\linewidth}}
\hline
\textbf{Step} & \textbf{Equation} \\
\hline

Initialization &
Previous belief distribution $[T]^{t-1} \sim \mathcal{N}([\bar{T}]^{t-1},P^{t-1}),$ where $[\bar{T}]^{t-1}$ is the mean and $P^{t-1}$ is the covariance.\\ &
$
[\bar{T}]^{0} = \mathbb{E} \left[[T]^{0}\right], \quad
P^{0} = \mathbb{E}\!\left[([T]^{0}-[\bar{T}]^{0})
([T]^{0}-[\bar{T}]^{0})^{\tp}\right]
$ \\

for $t \in \{1,...,\infty\}$
\\

Sigma points &
$
\mathcal{X}^{t-1} =
\Big[
[\bar{T}]^{t-1},
[\bar{T}]^{t-1}+\gamma\sqrt{P^{t-1}},
[\bar{T}]^{t-1}-\gamma\sqrt{P^{t-1}}
\Big]
$ \\

Time update (state propagation) &
$
\mathcal{X}_{*}^{t|t-1} = f(\mathcal{X}^{t-1},[\dot{E}_{gen}]^{t-1},[0])
$ \\

Predicted state mean &
$
[\bar{T}]^{t}_{(-)}
= \displaystyle\sum_{i=0}^{2n} \omega_i^{m}[\mathcal{X}_{*}^{t|t-1}]_i
$ \\

Predicted state covariance &
$
P^{t}_{(-)}
= \displaystyle\sum_{i=0}^{2n} \omega_i^{c}
([\mathcal{X}_{*}^{t|t-1}]_i-[\bar{T}]^{t}_{(-)})
([\mathcal{X}_{*}^{t|t-1}]_i-[\bar{T}]^{t}_{(-)})^{\tp}
+ R^{w}
$ \\

Measurement propagation &
$
[\hat{T}_\text{obs}]^{t}_{(-)} = H [\bar{T}]^t_{(-)}
$ \\

Innovation covariance &
$
{P}^{yy}
= HP^t_{(-)}H^{\tp} + {R}^{v}
$ \\

Cross covariance &
$
{P}^{xy}
= P^t_{(-)}H^{\tp}
$ \\

Kalman gain &
$
{K}^{t} = {P}^{xy}
({P}^{yy})^{-1}
$ \\

State update &
$
[\bar{T}]^{t} =
[\bar{T}]_{(-)}^{t}
+ {K}^{t}([T_{\text{obs}}]^{t}-[\hat{T}_\text{obs}]_{(-)}^{t})
$ \\

Covariance update (Joseph Form)&
$
{P}^{t}
=
({I} - {K}^{t} {H})\,
{P}^t_{(-)}\,
({I} - {K}^{t} {H})^{\tp}
+
{K}^{t} {R}^{w} ({K}^{t})^{\tp}
$ \\
\hline
\end{tabular}
\caption{Unscented Kalman Filter Equations with Additive Noise}\label{UKF Table}
\end{table}

The Unscented Kalman Filter in conjunction with our derived thermal model allows us to formally define our MPC procedure. As outlined by Algorithm \ref{MPC Algorithm}, the UKF first estimates the full state from the partial and noisy observations. 
With the estimated state as the initial condition (i.e., $[{T}]^{t} = [\hat{T}]^{t}$), we deterministically predict the future states over a horizon $h$ with our nonlinear thermal model given by Eq.~\eqref{thermal model}, and minimize Eq.~\eqref{cost function} with respect to the sequence of inputs $[\dot{E}_{gen}]^{0|t}\, \ldots,[\dot{E}_{gen}]^{h|t}$. 
The first input from the sequence is then applied to the system according to $[\dot{E}_{gen}]^{t} = [\dot{E}_{gen}]^{0|t}$, before the process is repeated at time $t+1$.

A few comments regarding our solution procedure are in order before discussing our results.
It is important to note that our MPC procedure minimizes the deterministic cost Eq.~\eqref{cost function} when the state is not fully observed and is instead represented by a belief distribution $[T]^t \sim \mathcal{N}([\bar T]^t, P^t)$ produced and updated by the UKF.
In other words, because the UKF models the state as a random variable, Eq.~\eqref{cost function} is not well-defined in the strict sense.
It is therefore common to consider minimizing the expected cost $\mathbb{E}[J|[\mathcal{I}]^t]$, where $J$ is the cost function in Eq.~\eqref{cost function}, over the sequence $[\dot{E}_{gen}]^{0|t},\ldots,[\dot{E}_{gen}]^{h|t}$, with $[\mathcal{I}]^t$ denoting the information state at time $t$ and the expectation taken over all future temperature sequences given $[\mathcal{I}]^t$.
The key difference between minimizing the expected cost and the deterministic cost Eq.~\eqref{cost function} arises from the coupling between controller and estimator performance in nonlinear systems such as ours.
Specifically, when minimizing the expected cost, the temperature belief covariance $P^t$ appears explicitly in the objective, introducing a coupling between the quality of the state estimate and the performance of the controller obtained via MPC.
Jointly designing the controller and estimator to account for this coupling is a challenging task in nonlinear systems, and we therefore fix the estimator design and discard the controller-estimator coupling by minimizing Eq.~\eqref{cost function} in place of the expected cost resulting in a separated controller-estimator design~\cite{findeisen2003state}.
Approaches that retain controller-estimator coupling exist in the literature~\cite{yan2005incorporating,bradford2017stochastic} but introduce additional complexity when implemented in a predictive control framework.
Accordingly, we do not propagate the state covariance or belief distribution within our predictive control algorithm.
Once the state estimate at time $t$ has been produced by the UKF, we set $[T]^{0|t} = [\hat T]^t$ and treat the optimization as a deterministic MPC problem, where the deterministic dynamical system Eq.~\eqref{thermal model} corresponds to the stochastic system Eq.~\eqref{noise thermal model} with the process noise sequence set to its mean, $\{[w]^{0|t},\ldots,[w]^{h|t}\} = \{0,\ldots,0\}$, which follows from $\mathbb{E}[[w]^{l|t}] = 0$ for all $l \in \{0,\ldots,h\}$.
The practice of substituting a state estimate into a deterministic optimal control problem is known as certainty equivalence (CE) control~\cite{haimovich2003optimality,kao2012directed}, and has been applied across a range of applications, yielding suboptimal but computationally efficient control for nonlinear systems~\cite{haimovich2003optimality,kulcsar2006optimal}.

\begin{algorithm}[ht!]
\caption{MPC Procedure}\label{MPC Algorithm}
\begin{algorithmic}[1]
\REQUIRE Thermal model mapping $[\dot{E}_{gen}]^{t-1}$ and $[T]^{t-1}$ to $[T]^{t}$ (Eq.~\eqref{thermal model}).
\FOR{$t = 0:\Delta t:\text{endTime}$}
    \STATE Estimate the current temperature $[\hat{T}]^{t}$ with the UKF (see Table~\ref{UKF Table}).
    \STATE Set the initial condition $[{T}]^{0|t} = [\hat{T}]^{t}$
    \STATE Deterministically predict future temperatures $[{T}]^{1|t}, \dots, [{T}]^{h+1|t}$ as a function of $[\dot{E}_{gen}]^{l|t}$ for $l \in\{0,\ldots,h\}$ with the thermal model.
    \STATE Solve the following optimization problem subject to~\eqref{thermal model} and~\eqref{eq:MPCineqCons1}:
    \begin{align}
\min_{[\dot{E}_{gen}]^{0|t},\ldots,[\dot{E}_{gen}]^{h|t}} &
\sum_{l=0}^{h}
\left([{T}]^{l|t}-[T_{ref}]\right)^{\tp}
Q
\left([{T}]^{l|t}-[T_{ref}]\right)
\nonumber\\
&+
\left([\dot{E}_{gen}]^{l|t}\right)^{\tp}
R
\left([\dot{E}_{gen}]^{l|t}\right)
\nonumber\\
&+
\left(
[\dot{E}_{gen}]^{l|t}
-
[\dot{E}_{gen}]^{l-1|t}
\right)^{\tp}
R_{\Delta}
\left(
[\dot{E}_{gen}]^{l|t}
-
[\dot{E}_{gen}]^{l-1|t}
\right). \nonumber
\end{align}

    \STATE Apply only the first input $[\dot{E}_{gen}]^t = [\dot{E}_{gen}]^{0|t}$ to the system.
\ENDFOR
\end{algorithmic}
\end{algorithm}

\section{Results} \label{Results Section}
To demonstrate the capabilities of the thermal model and MPC algorithm outlined in Sections \ref{Thermal Modeling Section} and \ref{Model Predictive Control Section}, we consider a CubeSat carrying an optical system in a noon-midnight orbit about the Earth (Fig.\ref{fig:noon-midnight-orbit}). 
A noon-midnight orbit was chosen to showcase the MPC algorithms ability to react in advance to known thermal disturbances, specifically entering or leaving Earths shadow and to meet the sub-Kelvin temperature requirement for wavefront stability. The orbit parameters are, inclination = 78.8$^\circ$, Right Ascension of ascending node = 0$^\circ$, argument of periapsis = 270$^\circ$, eccentricity = 0, period = 5677 s, and an altitude of 500 km.

\begin{figure}[b!]
	\centering
	\includegraphics[width=.7\textwidth]{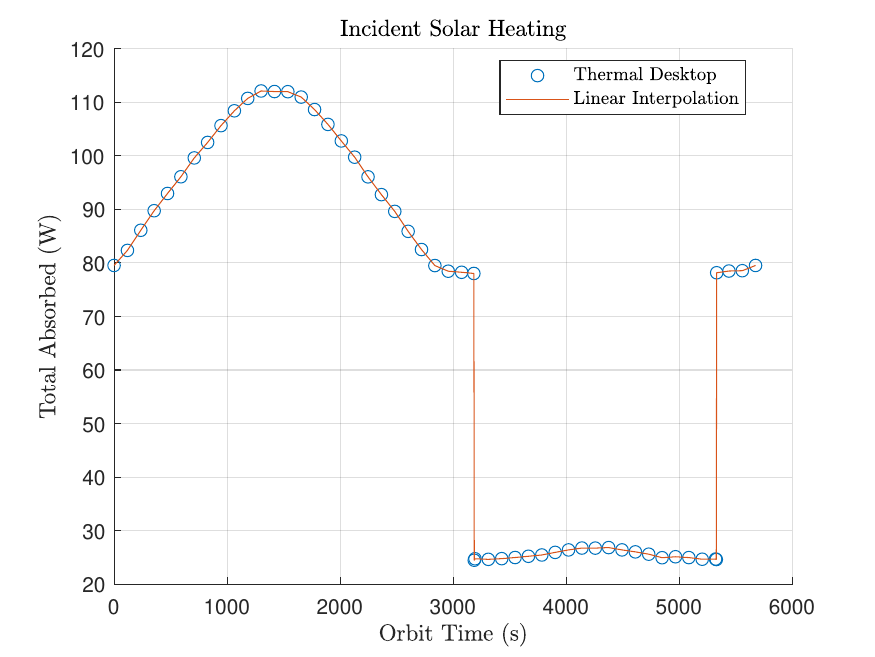}
	\caption{Total absorbed radiative heating from Solar, Albedo, and Earth infrared sources that is incident on the CubeSat's structure.}
	\label{orbital heating}
\end{figure}

The orientation of the CubeSat is such that the +z axis is always pointing toward the center of the Earth, which causes the CubeSats projected area to change relative to the sun. The solar, albedo, and Earth IR radiation absorbed by the CubeSat was determined in Thermal Desktop and is shown in Fig. \ref{orbital heating}. The large drop in absorbed radiation is due to the CubeSat entering the Earth's shadow which in turn blocks the incident solar radiation. The absorbed radiation does not drop to zero since Albedo and Earth IR still contribute while the CubeSat is in the Earths shadow.  
The CubeSat is assumed to be made of aluminum and have a surface emissivity of 0.5 for these results, although our approach is valid irrespective of the material selected.

\subsection{Sixty-Four Node In-Orbit Optical System Control and Simulation}
\begin{figure}[t!]
	\centering
	\includegraphics[width=1\textwidth]{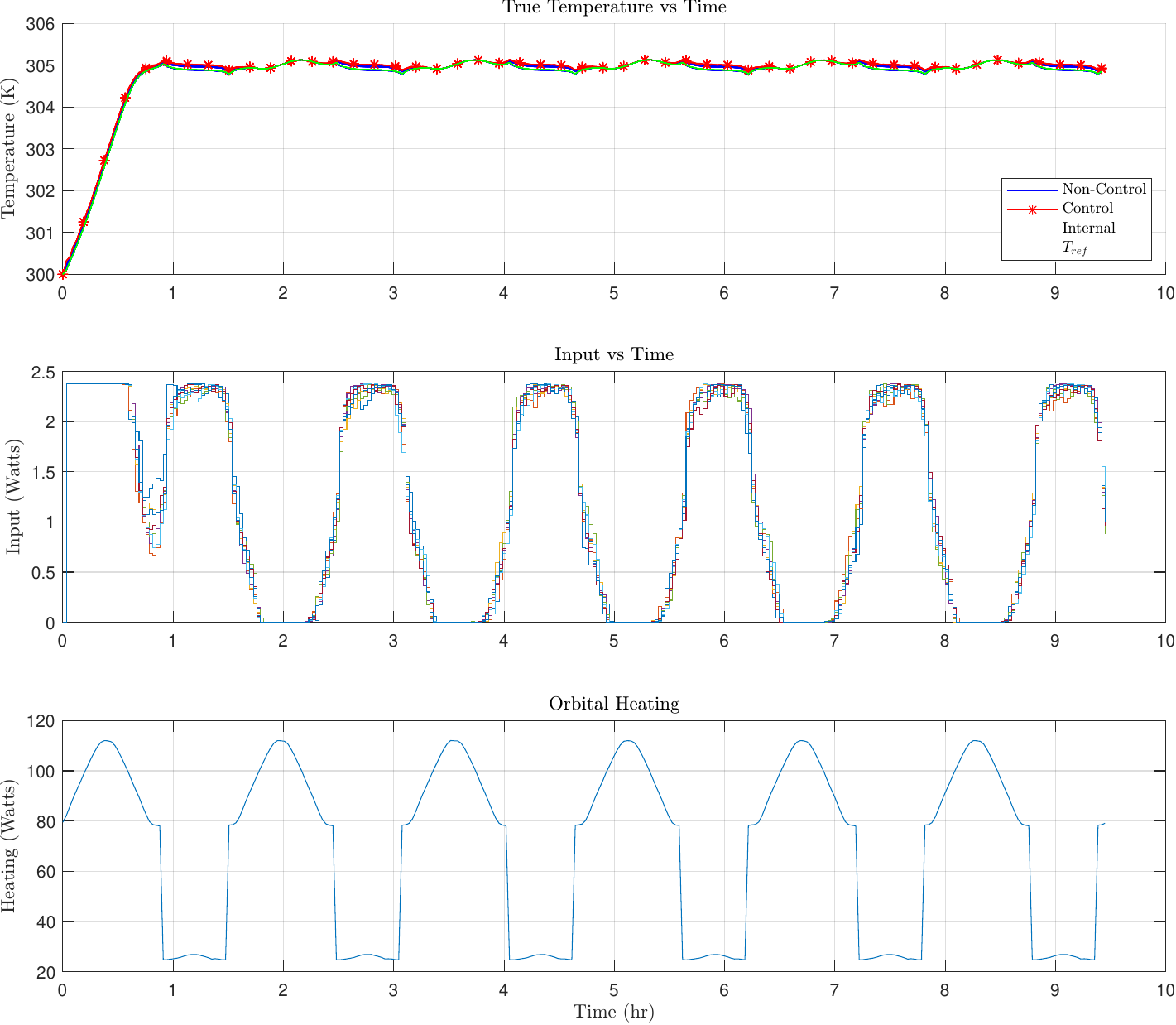}
	\caption{Sixty-four node MPC illustrating the true temperature and input responses on top of the orbital heating. The CubeSat optical system is initially at \(300\) \textmd{K}, but it is desired to maintain it at \(T_{ref} = 305\) \textmd{K}. The MPC algorithm calculates the necessary input \(\dot{E}_{gen}\) for the eight control nodes over a receding horizon at each time step to maintain the reference temperature. Each input is represented by a different color.}
	\label{64 Node MPC}
\end{figure}

To demonstrate the developed Model Predictive Control approach, we consider a CubeSat model consisting of sixty-four nodes for the optics and a sixty-fifth node to represent the CubeSats thin shell structure which encases the optics. 
We believe this to be a representative number of nodes since existing thermal control methods focus on controlling specific thermal zones \cite{Brooks2022}. In this application each of the sixty-four nodes and their control volumes can be thought as a thermal zone.
Due to the location of each node in the optical system, it is beneficial to define three types of nodes: internal, non-control, and control nodes. Internal nodes do not have an exposed surface and therefore cannot have a heater attached to them. Additionally, internal nodes do not experience radiation since they do not have an exposed surface. 
Non-control nodes have an exposed surface but do not have a heater attached to them, and finally control nodes have an exposed surface and a heater attached to them. 
These control nodes are located on the corners of the optical payload and are the only nodes used to control and measure the temperature of the optical system. 
Applying the MPC algorithm to control the temperature of the $n=64$ optical system results in Fig.~\ref{64 Node MPC}, where there are 8 control nodes, 8 interior nodes, and 48 non-control nodes. 

In the sixty-four node temperature response, we see that the CubeSat tracks the reference temperature with sub-Kelvin accuracy. Specifically, the max error for the estimated temperature is $0.40$ K with a mean error of $0.025$ K. The shape of this temperature response is dictated by the sizing of the weighting matrices, $Q$, $R$, and $R_\Delta$ in Eq.~\eqref{cost function}. 
If the weights contained in $Q$ were to be decreased, then the CubeSat's temperature would oscillate around the reference temperature leading to worse tracking. 
A decrease in $R$ would lead to less aggressive inputs, and a decrease in $R_\Delta$ would lead to faster changing inputs that could overreact to measurements.

\begin{figure}[t!]
	\centering
	\includegraphics[width=0.8\textwidth]{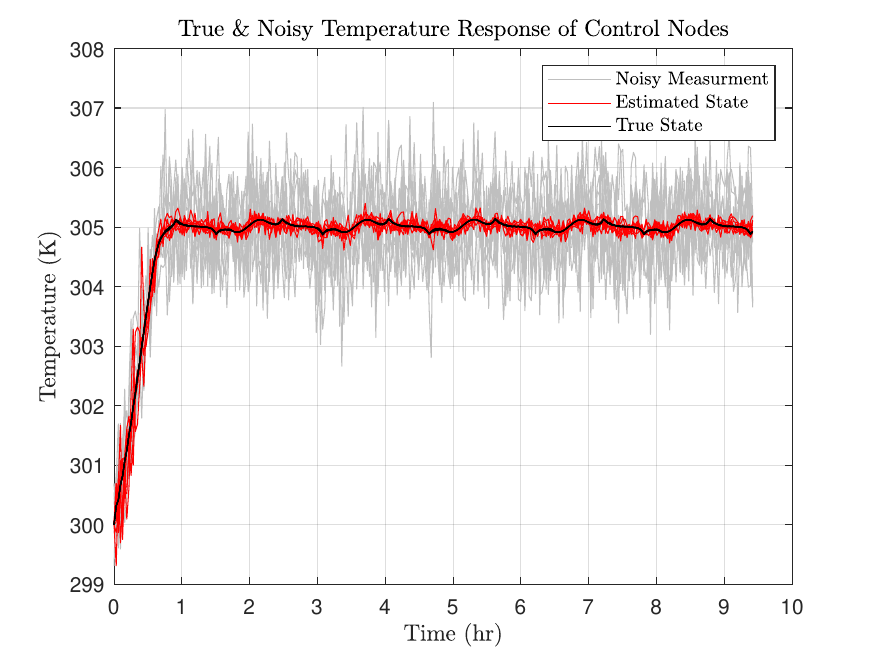}
	\caption{Control node temperature response for $n=64$ illustrating the effect of the Unscented Kalman Filter. The UKF processes the noisy partial measurement (gray) for each of the control nodes and estimates the true full state (red). The estimate of the true temperature is used in the MPC algorithm to determine the optimal input which is then applied to the real system (black).}
	\label{True and Noisy}
\end{figure}

One of the major benefits to employing MPC is the ability to control multiple heaters within a single framework, unlike PID-based methods. 
The ability to simultaneously design the operation of multiple heaters allows for gradient control in addition to temperature control, the former of which is difficult with classical methods \cite{vu2010independent}.
Temperature gradient control is especially important for optical component safety, as thermal gradients can result in unintended warping and lead to various optical aberrations \cite{spitzer1967thermal}.
Another important observation in Fig~\ref{64 Node MPC} is that the temperature of interior nodes lag behind the control nodes by one time step. 
This lag is due to the one time-step needed to propagate energy from one node to another and can be seen in the temperature response. Mainly at the start of the simulation where the interior node temperature (green) is slightly more to the right of the control node temperature (red) which indicates a delayed response for the interior nodes.

The control of the CubeSat thermal environment is subject to input constraints as unlimited heating (energy) is not possible in-orbit.
The input constraints imposed on our system are given by
\begin{align*}
    0\leq &[\dot{E}_{gen}]^{l|t} \leq 2.40 W, \quad l\in\{0,\ldots,h\},
\end{align*}
together with $[\dot{E}_{gen}]^{l|t}_{internal} = 0 W$ and $[\dot{E}_{gen}]^{l|t}_{NonControl} = 0 W$, since these nodes do not have attached heaters. 
The constraint value of $2.40~\text{W}$ was obtained by analyzing the power available from the satellite's solar panels, given by $P = \eta A G_s D_s$, where $\eta$ is the solar panel efficiency, $A$ is the projected surface area, $G_s$ is the solar irradiance, and $D_s$ is a duty cycle term representing the fraction of time the panels are exposed to the sun. Taking $\eta = 0.3$, $A = 0.078~\text{m}^2$ (corresponding to a single $22.63~\text{cm} \times 34.6~\text{cm}$ CubeSat face), $G_s = 1354~\text{W}/\text{m}^2$, and $D_s = 0.6$ yields $P = 19.1~\text{W}$, which, divided among the eight available heaters, gives $2.4~\text{W}$ per heater.

To emphasize the importance of the Unscented Kalman Filter to our approach, we compare the noisy, estimated, and true temperatures in Fig.~\ref{True and Noisy}. 
As seen in Fig.~\ref{True and Noisy},
the Unscented Kalman Filter removes the measurement noise and provides an estimate of the true temperature of the system at each internal node. 
The estimated temperature is used to determine the optimal input by minimizing the cost function Eq.~\eqref{cost function} using the procedure described in Algorithm~\ref{MPC Algorithm}. 
The optimal input is then applied to the true system. 
Note that, without the UKF, we would only have access to the noisy partial state, and would be unable to verify that the optical system remains in its strict temperature range.
Additionally, our control would be more susceptive to the measurement noise.

\begin{figure}[t!]
\centering
\includegraphics[width=1\textwidth]{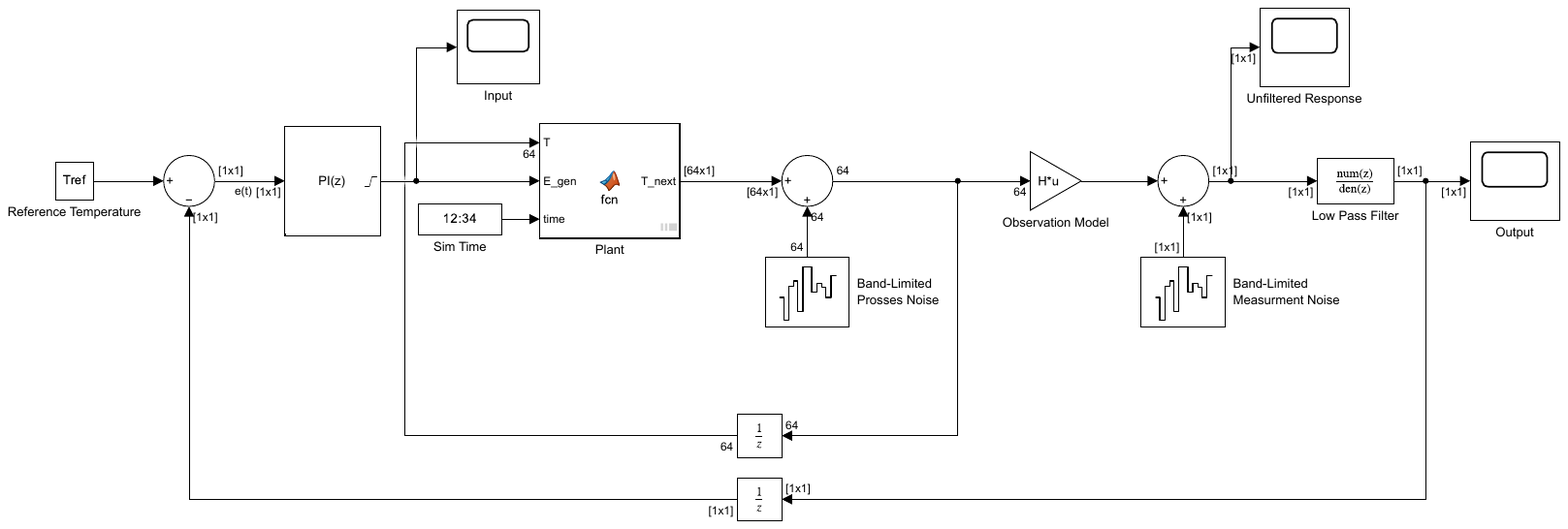}
\caption{The CubeSat temperature dynamics Eq.~\eqref{thermal model} are represented by the `Plant' block and are a function of the CubeSat's temperature, input, and orbital position. Gaussian zero mean noise is added to the process and the measurement to mimic real world errors in the process and sensor. $H$ is the linear measurement model and is used to read the temperature of node one. A standard first order low-pass filter is used to remove high frequency noise that would, if left unfiltered, lead to ineffective error calculations.  The output of the low-pass filter is compared to the reference temperature to generate the error signal $e(t)$. The error signal combined with the PI control structure determines the next input to regulate the CubeSat's temperature.}
\label{PID Control}
\end{figure}

To illustrate the benefit of the MPC algorithm compared to traditional methods, we compare our control methodology to that of PID. 
Recall that PID does not require an explicit model, reacts to disturbances once they have effected the systems output, does not provide a optimal control to correct the output, and has difficulty with complicated nonlinear systems \cite{ziegler1942optimum,aastrom2007feedback,vrabel2025optimum}.
In this comparison, we designed a proportional-integral (PI) controller to control the temperature of the sixty-four node CubeSat which is treated as a black box. 
PI control is selected for comparison as it is considered a leading solution in thermal control~\cite{Brooks2022}.
Since PID is single-input single-output (SISO), only the first node is used to sense and control the temperature of the CubeSat. Additionally, input saturation defined as $0 W \leq \dot{E}_{gen} \leq 19.1 W$ and anti-windup for the integral gain were included to prevent integration of the error during input saturation~\cite{da2018analysis}. The block diagram illustrating the PID control is shown in Fig.~\ref{PID Control}. The flow of the block diagram is as follows.
At each time step, the current temperature $[T]^t \in \mathbb{R}^{64 \times 1}$ and input $\dot{E}_{gen} \in \mathbb{R}^{1\times 1}$ are used to compute the next temperature $[T]^{t+1}$ via the black-box `Plant' model, to which Gaussian process noise is added to simulate plant errors. 
The resulting temperature is then passed through the measurement model $H$ to obtain the temperature of node one, corrupted by Gaussian measurement noise to simulate real-world sensor errors, and subsequently filtered by a first-order low-pass filter to attenuate high-frequency noise. 
The filtered measurement is compared against the reference temperature to form the error signal $e(t)$, which is then passed to a discrete-time PI compensator to determine the next control input $\dot{E}_{gen}$ for regulation of the optical component temperature.

\begin{figure}[t!]
\centering
\includegraphics[width=0.95\textwidth]{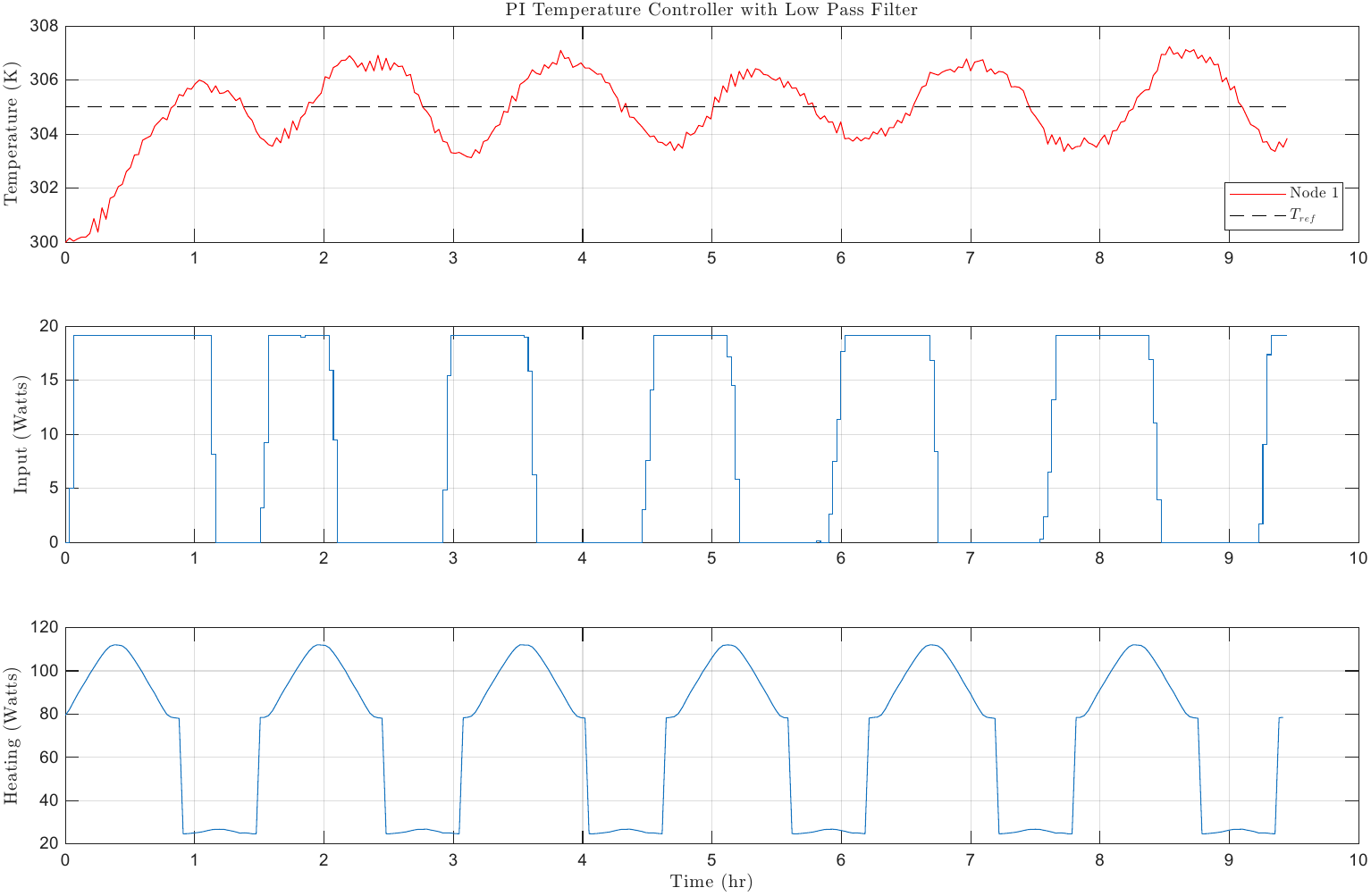}
\caption{Sixty-Four node PI Control illustrating its temperature tracking capability and its input responses on top of the orbital heating. The CubeSat optical system is initially at \(300\) \textmd{K}, but it is desired to maintain it at \(T_{ref} = 305\) \textmd{K}. The PI compensator calculates the necessary input from the error signal $e(t)$ to track the reference temperature.}
\label{PID Response}
\end{figure}

Implementing the PI control described by Fig.~\ref{PID Control} results in Fig.~\ref{PID Response}. 
Moderate thermal control is achieved with the PI control methodology, however, the absence of predictive control is clear. 
The input response is not out of phase with the external orbital heating and the input only `turns on' when the temperature of the CubeSat is below the reference temperature. 
These trends are not found when predictive control is used, which can be observed in Fig. \ref{64 Node MPC}. The $90 ^\circ$ phase shift between the orbital heating and heater inputs results from the MPC algorithm's ability to anticipate future orbital heating disturbances. By incorporating predicted disturbances into the optimization process, the controller applies heater inputs in advance, producing the observed phase shift. Additionally, due to the PID frame work, only the temperature of one node is used for feedback and the other node temperatures are unknown. 
Since the majority of the optical payloads temperature is unknown, it is difficult to validate the optical safety. 
Moreover, thermal gradient control is not possible as only one heater can be used for control. 
These drawbacks are overcome with the MPC framework.

To quantitatively compare MPC and PID control, a temperature tracking and a power-per-orbit analysis is conducted. For the temperature tracking analysis we consider the root mean square error (RMSE) of node ones temperature around $T_{ref}$ for six orbit periods after the initial transients have settled.
We remove the first hour from our analysis to eliminate initial transients that do not correspond to steady-state tracking. 
For the orbit-power analysis we integrate the input response versus time and divide by the number of orbits which results in the average energy used per orbit. 
Next, we divided the average energy by the period of one orbit to obtain the average power per orbit. 
The results for these comparisons is displayed in Table~\ref{PID comp}.
\begin{table}[h]
\centering
\begin{tabular}{cccc}
\toprule
PID RMSE (K) & MPC RMSE (K) & PID Power/Orbit (W) & MPC Power/Orbit (W)  \\
\midrule
1.16 & 0.13 & 7.57 & 7.98 \\
\bottomrule
\end{tabular}
\caption{}
\label{PID comp}
\end{table}
From Table~\ref{PID comp} it is clear that the MPC provides considerably better tracking than the PID framework at the cost of slightly more power per orbit. 
Specifically, there is a 88\% reduction in the temperature RMSE and a 5\% power increase per orbit from PID to MPC. 
As expected, the MPC temperature RMSE is significantly lower than that of PID which can be accredited to the MPC cost function. 
The cost function described by Eq.~\eqref{cost function} is minimal when the heater input forces the CubeSat's optical temperature to be at the reference temperature. 
However, due to the increased temperature tracking, more heater power is used to satisfy the cost function resulting in slightly more power needed per orbit. 
For these reasons, MPC should be considered when precise gradient and temperature control is needed for spacecraft carrying sensitive components, as is the case for space-based telescopes \cite{gaudi2020habitable}.

Due to the PID's formulation, it is not possible to determine the temperature of the other nodes since this PID control does not employ an explicit model, instead modeling the system as a black box. 
However, we do know that node one does not meet the sub-Kelvin thermal requirement to provide the necessary wavefront stability, as the estimated max temperature error is $2.24$ K. The maximum temperature error is computed as the maximum absolute deviation of the estimated temperature from the $305$ K reference, taken over both space and time once the initial transients have subsided.
In the MPC formulation, we have access to the temperature estimates of all nodes within the optical system, yielding a maximum temperature error of $0.40$ K across the $64$ node system, which meets the sub-Kelvin thermal requirement. 
Furthermore, the mean error across the payload is $0.025~\text{K}$, computed by averaging the node temperatures spatially and temporally and comparing the result to the $305~\text{K}$ reference over an interval following the initial transient.
Finally, the temperature response exhibits a standard deviation of $0.079~\text{K}$, computed by first averaging the node temperatures at each time step and then taking the standard deviation of this spatially averaged signal over time.
This meets the requirements for many applications~\cite{knapp2020demonstrating,li2025design,song2025thermal}.
It should be noted that improved performance may be realized by optimization over the thermal properties (e.g., $\epsilon$, $\rho$, $k$, $c_p$).

\section{Conclusion}\label{sec:conclusion}

In this paper, we developed a nonlinear thermal model for a CubeSat carrying an optical payload in a Noon-Midnight orbit, representative of a general spacecraft with precision optics.
Maintaining a stable thermal environment is critical for the success of spacecraft carrying precision optics, as thermal effects such as optical misalignment and thermally-induced expansion can deteriorate optical performance and lead to poor science images~\cite{kristwavefrontcontrol}.
These effects can be prevented by regulating the temperature of the optical system to within its designed operating range.
External heating sources encountered during orbit, including solar, Earth IR, and albedo flux, directly affect the temperature of the optics and, if left unregulated, cannot be guaranteed to keep the optical system within its safety and performance limits.
To this end, we formulated an model predictive control (MPC) methodology to regulate the spacecraft optical system to its designed temperature by sensing the current thermal state, predicting future temperatures, and determining the optimal heating input at each time step.
Since only partial, noise-corrupted temperature measurements are available in practice, an Unscented Kalman Filter (UKF) is employed to estimate the full thermal state of the system, which is then used to initialize the MPC optimization under a certainty-equivalence formulation.
The developed thermal model was validated against industry-standard thermal software, and the full control and estimation framework was demonstrated on a sixty-four node CubeSat model in a realistic orbital environment.
A comparison between our MPC approach and the more traditional PID framework was conducted to illustrate the advantages of the proposed approach.
We demonstrated that our proposed MPC approach meets the sub-Kelvin thermal requirement for precision optical payloads and achieves a 88\% reduction in tracking error at the cost of slightly increased power consumption per orbit compared with PID. 

\section*{Acknowledgments}
Portions of this research were supported by funding from the Technology Research Initiative Fund (TRIF) of the Arizona Board of Regents and by generous anonymous philanthropic donations to the Steward Observatory of the College of Science at the University of Arizona. The authors also acknowledge that  assistance of AI tools were used in the preparation of this manuscript to improve readability, grammar, and language in this work.

\end{document}